\documentclass[%
 aip,
cp,  
 amsmath,amssymb,
 reprint,%
]{revtex4-2}

\usepackage{graphicx}
\usepackage{comment}
\usepackage{dcolumn}
\usepackage{bm}

\usepackage[utf8]{inputenc}
\usepackage[T1]{fontenc}
\usepackage{mathptmx} 

\begin{document}

\title{How `Nonvariational' Are Approximate Coupled Cluster Methods In Practice?}

\author{Jan M.L. Martin} 
 \email[Corresponding author: ]{gershom@weizmann.ac.il}
\author{Emmanouil Semidalas}%
 \email{emmanouil.semidalas@weizmann.ac.il}
\affiliation{
  Department of Molecular Chemistry and Materials Science, Weizmann Institute of Science, 7610001 Re\d{h}ovot, Israel.
}

\date{\today} 

\begin{abstract}
While limited coupled cluster theory is \textit{formally} nonvariational, it is not broadly appreciated whether this is a major issue \textit{in practice}. We carried out a detailed comparison with \textit{de facto} full CI energies for a relatively large and diverse set of molecules. Fully iterative limited CC methods such as CCSDT, CCSDTQ, CCSDTQ5 do represent practical upper bounds to the FCI energy. While quasiperturbative approaches such as CCSD(T) and especially CCSDT(Q) may significantly over-correlate molecules if there is significant static correlation, this is much less of an issue with  Lambda approaches such as CCSDT(Q)$\Lambda$.
\end{abstract}

\maketitle

\section{\label{sec:intro}Introduction and Statement of the problem}

Coupled cluster (CC) theory (for a comprehensive review, see Ref.\onlinecite{ShavittBartlett}) has become the tool of choice for accurate wavefunction theory (WFT) electronic structure calculations.

An untruncated coupled cluster \textit{ansatz}, $\Psi=\exp(\hat{T})\psi_0$, where the cluster operator $\hat{T}=\hat{T}_1+\hat{T}_2+\ldots+\hat{T}_n$, is merely a clumsy way of carrying out an FCI (full configuration interaction) calculation, which corresponds to the exact solution within a given finite basis set. However, limited CC, with a truncated cluster operator such as $\Psi_{\rm CCSD}=\exp(\hat{T}_1+\hat{T}_2)\psi_0$ and $\Psi_{\rm CCSDT}=\exp(\hat{T}_1+\hat{T}_2+\hat{T}_3)\psi_0$ not only converges much more rapidly to the FCI limit, but unlike limited CI is rigorously size extensive. The CCSD(T) method\cite{Rag89,Wat93}, in particular, is often referred to as ``the gold standard of quantum chemistry'' (following T. H. Dunning, Jr.), and is widely used as a reference or `sanity check' for low-cost methods like DFT (density functional theory).

On the flip side, limited CC methods are nonvariational and hence (again, unlike limited CI) does not yield a guaranteed upper bound to the system's total energy. Given that variational energies are an upper bound for the exact energy, and that recently\cite{Pollak2021,Pollak2023} there has been revived interest in establishing rigorous \textit{lower} limits for the exact energy, this offers the tantalizing prospect of a rigorous error bound on approximate energies.

There have been efforts in variational coupled cluster theory (e.g., Refs.\onlinecite{VCC1,VCC2,VCCvsUCC}), but very few researchers have adopted them.
Prof. Eli Pollak (Weizmann Institute of Science), at the Ph.D. defense of one of us (ES), wondered aloud why. This made the senior author ponder just how serious a practical issue (as distinct from a formal one) the nonvariational character of limited CC truly is. We shall address this question in the present brief contribution.

\section{\label{sec:methods}Computational methods}

Most calculations in this paper were carried out using the MRCC\cite{MRCC} general coupled cluster program of K\'allay and coworkers. Some additional calculations were performed using the public and development versions of the CFOUR program system\cite{CFOUR}. The lion's share of the raw data are already available online in the Supporting Information of Ref.\onlinecite{jmlm330}; the CCSDTQ567 energies for a subset of systems can be obtained upon request from the author.

The approximate CC methods considered include CCSD[T] (a.k.a., CCSD+T(CCSD))\cite{CCSDTpart1}; 
CCSD(T)\cite{Rag89,Wat93}; CCSDT-1a and CCSDT-1b\cite{CCSDT-1}; CCSDT-2\cite{CCSDT-3} and CCSDT-3\cite{CCSDT-3}; full CCSDT\cite{CCSDTpart1,CCSDTpart2}; CCSDTQ-1\cite{CCSDTQ-1} and CCSDTQ-3\cite{KallayGauss1}; full CCSDTQ\cite{CCSDTQ}; CCSDT(Q)\cite{mrcc8}; CCSD(T)$_\Lambda$\cite{lambdastanton1,lambdastanton2,lambdabartlett1,lambdabartlett2}, CCSDT(Q)$_\Lambda$\cite{KallayGauss1,KallayGauss2}, and CCSDTQ(5)$_\Lambda$\cite{KallayGauss1,KallayGauss2}; finally, higher-order fully iterative CC methods\cite{GeneralCC1,GeneralCC2} such as CCSDTQ5 and CCSDTQ56. It needs to be kept in mind that asymptotic CPU time scaling of fully iterative $m$-fold CC theory will be $O(n_{\rm occ}^mN_{\rm virt}^{m+2})N_{\rm iter}$ (where $n_{\rm occ}$ and $N_{\rm virt}$ represent the numbers of occupied and virtual orbitals, respectively, and $N_{\rm iter}$ is the number of iterations), 
compared to $O(n_{\rm occ}^{m-1}N_{\rm virt}^{m+1}N_{\rm iter})$
$+O(n_{\rm occ}^mN_{\rm virt}^{m+1})$ 
for a quasiperturbative `parentheses' method and (because of the need to solve for the $\Lambda$ `left eigenvector' in addition to the CC `right eigenvector') approximately $2\times O(n_{\rm occ}^{m-1}N_{\rm virt}^{m+1})N_{\rm iter}$+$O(n_{\rm occ}^mN_{\rm virt}^{m+1})$ for a $\Lambda$ approach. 

The molecules considered are the 140 species from the W4-11 thermochemical benchmark\cite{jmlm235} and its 96-member subset W4-08\cite{jmlm215}. These span a range of inorganic and organic molecules, first-row and second-row, and range from essentially purely dynamical correlation (such as H$_2$O and SiF$_4$) to strong static correlation (such as O$_3$, S$_4$, C$_2$, and BN).

Basis sets considered are the Dunning correlation consistent\cite{Dunning1989,Woon1993} basis sets. Specifically, we considered cc-pVDZ (correlation consistent polarized double zeta) with the polarization function on hydrogen removed --- which we denoted cc-pVDZ(d,s) --- and cc-pVDZ with all polarization/angular correlation functions removed, which we denoted cc-pVDZ(p,s).

As we showed almost two decades ago\cite{jmlm205}, high-order coupled cluster increments, such as $E[T_5]=E[{\rm CCSDTQ5}] - E[{\rm CCSDTQ}]$, converge very rapidly with the basis set, in fact the more rapidly so as the excitation level increases (and increasingly, static rather than dynamical correlation is sampled). This has been exploited in high-accuracy computational thermochemistry protocols such as W4 theory\cite{jmlm200,jmlm269} and HEAT\cite{HEAT,HEAT2,HEAT3,HEAT4}. Hence, for the reference correlation energies, we approximated CCSDTQ56/cc-pVDZ(d,s) reference energies as follows:
\begin{itemize}
    \item where CCSDTQ5/cc-pVDZ(d,s) is available, E[CCSDTQ5/cc-pVDZ(d,s)]+E[CCSDTQ56/cc-pVDZ(p,s)]-E[CCSDTQ5/cc-pVDZ(p,s)]
    \item otherwise, where CCSDTQ(5)$_\Lambda$/cc-pVDZ(d,s) is available, E[CCSDTQ5/cc-pVDZ(d,s)]+E[CCSDTQ5(6)$_\Lambda$/cc-pVDZ(p,s)]-E[CCSDTQ(5)$_\Lambda$/cc-pVDZ(p,s)]
    \item for the largest species, E[CCSDTQ/cc-pVDZ(d,s)]+E[CCSDTQ5(5)$_\Lambda$/cc-pVDZ(p,s)]-E[CCSDTQ/cc-pVDZ(p,s)]
\end{itemize}

\section{\label{sec:results}Results and Discussion}

\begin{figure}
\includegraphics[width=12cm,trim={0 0cm 0 0},clip]{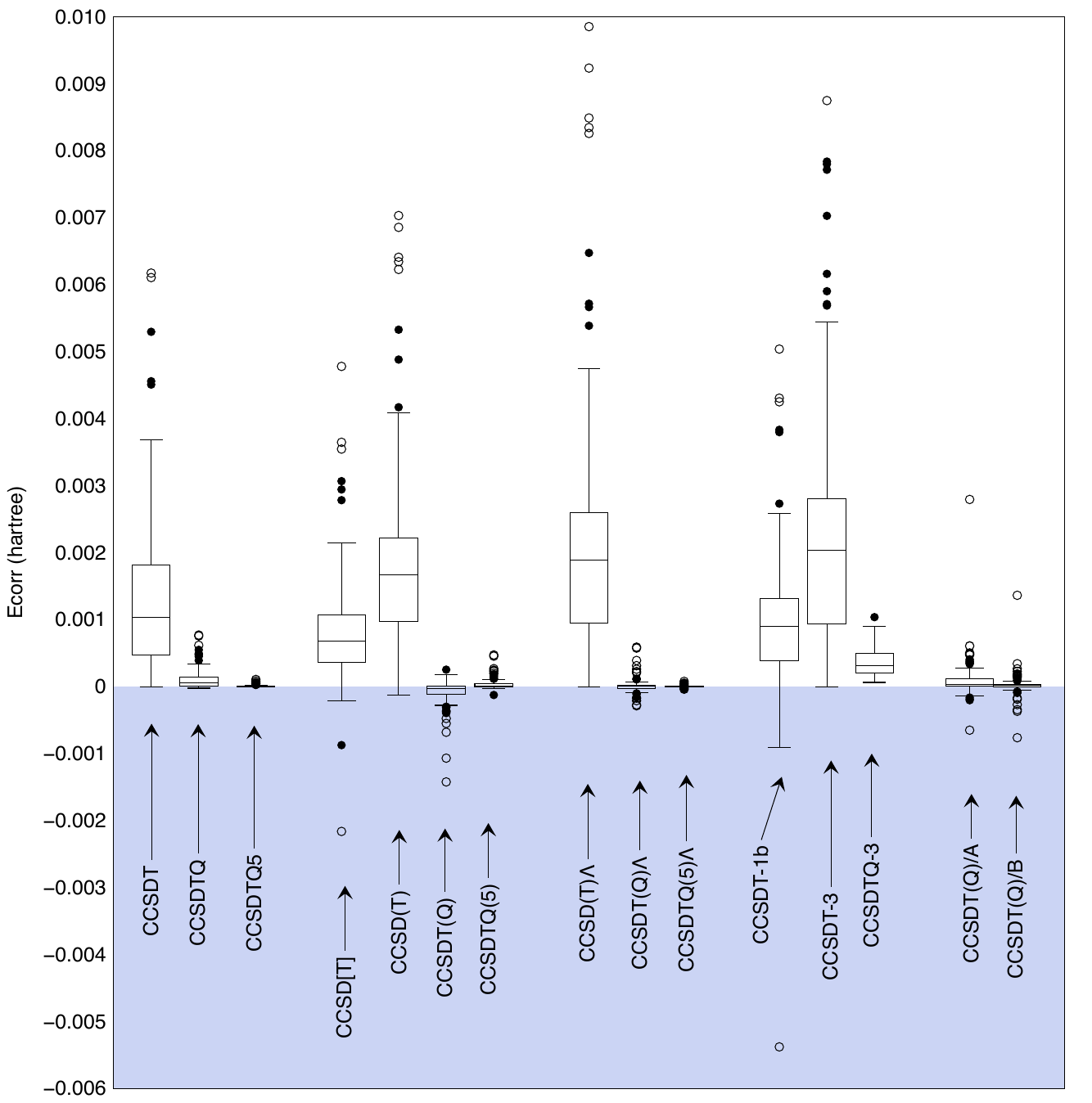}
\caption{\label{fig:epsart} Box plot of errors in W4-08 correlation energies (hartree) relative to the \textit{de facto} FCI reference data. The cc-pVDZ(d,s) basis set was used throughout. The `blue sea' indicates energies that violate the variational criterion. The box encompasses the middle half of the distribution, i.e., the IQR (interquartile range) between the 25th and 75th percentile (Q1 and Q3, respectively). Whiskers span from Q1-1.5IQR to Q3+1.5IQR. Filled circles are outliers, open circles are `extreme outliers', $<$Q1-3IQR or $>$Q3+3IQR.}
\end{figure}

We first attempted to ensure, for a subset of molecules, that the fully iterative CCSDTQ56 was close enough to full CI. In order to do so, we carried out fully iterative CCSDT567 calculations and assessed $E[{\rm CCSDTQ567}]-E[{\rm CCSDTQ56}]$. The largest correlation energy increments from connected septuple (!) excitations were seen for such troublesome species as C$_2$ and BN, and even there did not exceed several \textit{microhartree}; the largest contribution found was 6 $\mu E_h$ for ozone. For the remaining species, connected septuples contributions were on the other of 1 $\mu E_h$ or less, meaning that CCSDTQ56 is effectively full CI quality.

A box-and-whiskers plot of errors for various approximate coupled cluster methods is given in Figure~\ref{fig:epsart}. First of all, the fully iterative approaches appear to be variational in practice, albeit not formally. Also, as can reasonably be expected, CCSDTQ5 is extremely close to the FCI limit: the largest connected sextuples contribution is 85 $\mu E_h$ for ozone (0.053 kcal/mol). 
The largest connected quintuples contribution is rather more substantial, skirting the edges of a \textit{milli}hartree for some systems: 0.77 m$E_h$ (0.49 kcal/mol) for S$_4$, followed by 0.67 m$E_h$ (0.42 kcal/mol) for ozone. That connected quadruples are quite important (e.g., on the order of 6 m$E_h$, or 3.8 kcal/mol for molecules like S$_4$, ozone, and FOOF) is well-known by now (e.g., Ref.\onlinecite{jmlm200}).

The oldest `parenthetical' method, CCSD[T] a.k.a. CCSD+T(CCSD), predictably overshoots the correlation energy in a large subset of cases. This is mostly remedied in the  familiar `gold standard' CCSD(T) method, which includes a usually repulsive fifth-order $E^{[5]}_{ST}$-like term,\footnote{For detailed analyses of coupled cluster contributions in terms of M{\o}ller-Plesset perturbation theory, see Refs.\onlinecite{Kucharski1986} and \onlinecite{Cremer2001}} where in our sample only singlet BN diatomic has a CCSD(T) correlation energy that dips below the FCI limit. It is well established\cite{HEAT,jmlm173,jmlm200,jmlm205} that the generally good performance of CCSD(T) for thermochemistry is a consequence of felicitous error compensation between neglect of higher-order triples (typically repulsive) and of connected quadruples (universally attractive).
The CCSDT(Q) method is much closer to the FCI limit, but can be seen here to exceed the FCI correlation energy by up to 1.5 millihartree for a number of molecules — not just the usual suspects like BN, C$_2$, O$_3$, and S$_4$, but also N$_2$O and a number of others. Even for CCSDTQ(5), there is still a degree of nonvariational character.

Let us now turn to the `lambda coupled cluster' approaches\cite{lambdastanton1,lambdastanton2}. CCSD(T)$_\Lambda$ stays above FCI for all molecules, but is substantially further away from FCI than CCSD(T) for molecules with strong static correlation. CCSDT(Q)$_\Lambda$, on the other hand, represents an unqualified improvement over CCSDT(Q) -- in fact, its error distribution looks more akin to CCSDTQ(5) than to CCSDT(Q). The largest `nonvariational outlier' of CCSDT(Q)$_\Lambda$ is C$_2$, at 0.23 millihartree --- nearly an order of magnitude less than for CCSDT(Q). It would not be unreasonable to say that CCSDT(Q)$_\Lambda$ otherwise compares favorably to fully iterative CCSDTQ, at lesser cost. Ratcheting the CC$_\Lambda$ expansion one notch up, CCSDTQ(5)$_\Lambda$ has a very narrow error distribution like the much more expensive, fully iterative, CCSDTQ5 --- but unlike the latter, does have about one-quarter of the distribution below FCI. This is less serious than it sounds, however, considering that the worst case, singlet BN, is overcorrelated by just 61 microhartree (0.038 kcal/mol). Finally, CCSDT5(6)$_\Lambda$ (not displayed in Figure 1) does an exceedingly good job of capturing the connected sextuples: the largest discrepancy with fully iterative CCSDTQ56 is 5 microhartree for the pathologically multireference BN diatomic.

Turning now to the approximate iterative approaches: CCSDT-1b has a narrower box (encompassing 50\% of data points) than CCSD(T), but a much worse overcorrelation problem. CCSDT-3, on the other hand, has broader boxes than both CCSDT and CCSD(T), but nowhere dips below FCI. The latter is also true for CCSDTQ-3, which has broader boxes than CCSDT(Q) or CCSDT(Q)$_\Lambda$. The CCSDT-3 spread is noticeably wider than for full CCSDT: this reflects the thermochemical importance of $T_3-T_3$ coupling terms, which are present in CCSDT but missing in CCSDT-3\cite{Cremer2001}, and start at fifth order with $E^{[5]}_{TT}$. Note that CCSDT-1b and CCSDT-3 both scale as $O(n_{\rm occ}^3N_{\rm virt}^4N_{\rm iter})$ rather than the $O(n_{\rm occ}^3N_{\rm virt}^5N_{\rm iter})$ of CCSDT.

Finally, at the far right of Figure~\ref{fig:epsart} are the CCSDT(Q)$_{/A}$ and CCSDT(Q)$_{/B}$ approximations\cite{KallayGauss2} to CCSDT(Q)$_\Lambda$. CCSDT(Q)$_{/A}$ is clearly a poor substitute, while CCSDT(Q)$_{/B}$ appears to be preferable to CCSDT(Q) but remains inferior to CCSDT(Q)$_\Lambda$.

\section{Conclusion}

While limited coupled cluster theory is \textit{formally} nonvariational, it is not broadly appreciated whether this is a major issue \textit{in practice}. Through comparison with \textit{de facto} full CI energies for a relatively large and diverse set of molecules, we were able to establish that:
\begin{itemize}
    \item Fully iterative limited CC methods such as CCSDT, CCSDTQ, CCSDTQ5 do represent upper bounds to the FCI energy in practice.
    \item In contrast, quasiperturbative approaches such as CCSD(T), and especially CCSDT(Q), may significantly over-correlate molecules if there is significant static correlation.
    \item Lambda coupled cluster methods such as CCSDT(Q)$\Lambda$ and CCSDTQ(5)$_\Lambda$ strongly mitigate the issue.
    \item Fully iterative CCSDTQ56 as well as CCSDTQ5(6)$\Lambda$ are for all intents and purposes of full CI quality.
\end{itemize}

\begin{acknowledgments}
The authors would like to acknowledge Prof. Eli Pollak (Weizmann Institute of Science) for raising the issue, and Prof. John F. Stanton (Quantum Theory Project, University of Florida) for helpful discussions. This research was funded by the Israel Science Foundation (grant 1969/20) and by the Uriel Arnon Memorial Research Fund for Artificial Intelligence in Materials Science. The authors would like to thank Dr. Margarita Shepelenko for critical reading of the draft.
\end{acknowledgments}

\bibliography{nonvarCC}

\end{document}